\documentclass[12pt]{article}
\usepackage{amssymb}
\def\h#1{\hat #1}
\def\p{\partial}
\def\t#1{\tilde #1}
\def\ad{\mbox{ad}\,}
\def\b#1{{\mathbb #1}}

\def\c#1{{\cal #1}}

\def\Dirac{{\raise0.09em\hbox{/}}\kern-0.69em D}

\def\kbar{{\mathchar'26\mkern-9muk}}  

\def\t#1{\tilde #1}
\def\tr{\mbox{Tr}}

\def\wm{\mathbin{*}}
\def\hwm{\mathbin{\hat *}}
\def\k{\kern-.1em\mathbin{,}\kern-.1em}
\def\hk{\kern.12em\raise-1em\hbox{$\hat{\raise1em\hbox{,}}$}\kern.12em}
\newcommand{\initiate}{\setcounter{equation}{0}}
\begin{document} 

\title{External Fields as Intrinsic Geometry}

\author{J. Madore,$\strut^{1,2}$\  S. Schraml,$\strut^{2,3}$\ 
        P. Schupp$\strut^{3}$\ and J. Wess$\strut^{2,3}$\\[15pt]
        $\strut^{1}$Laboratoire de Physique Th\'eorique\\
        Universit\'e de Paris-Sud, B\^atiment 211, F-91405 Orsay
\and    $\strut^{2}$Max-Planck-Institut f\"ur Physik\\
        F\"ohringer Ring 6, D-80805 M\"unchen
\and    $\strut^{3}$Sektion Physik, Ludwig-Maximilian Universit\"at\\
        Theresienstra\ss e 37, D-80333 M\"unchen}

\date{}

\maketitle

\begin{abstract}
There is an interesting dichotomy between a space-time
metric considered as external field in a flat background and the
same considered as an intrinsic part of the geometry of space-time. 
We shall describe and compare two other external
fields which can be absorbed into an appropriate redefinition of the
geometry, this time a noncommutative one. We shall also recall some
previous incidences of the same phenomena involving bosonic field
theories. It is known that some such theories on the commutative
geometry of space-time can be re-expressed as abelian-gauge theory
in an appropriate noncommutative geometry. The noncommutative
structure can be considered as containing extra modes all of whose
dynamics are given by the one abelian action. 
\end{abstract}

\vfill
\noindent
LMU-TPW 00-00\\
\noindent
MPI-PhT/00-00\\
\noindent
DRAFT Version CVS:1.17
\newpage

\initiate
\section{Introduction and motivation}

It is known that some bosonic field theories on the commutative
geometry of space-time can be re-expressed as abelian-gauge theory in
an appropriate noncommutative geometry. This fact is quite the
analogue of the dichotomy in general relativity between the components
of a metric considered as external fields in a flat background and the
same components considered as defining the metric and therefore a
non-flat geometry. In the next section we mention very briefly a
certain number of examples which have been considered in the past and
which exhibit the property of an external field which can be
incorporated into a redefinition of the basic geometry.  The
noncommutative structure can be considered as containing extra modes
all of whose dynamics are given by the one abelian action. An example
is afforded by the Yang-Mills-Higgs-Kibble action of the standard
model~\cite{DubKerMad89a,ConLot90}.  Somewhat analogous results are also
known, for example, for non-relativistic hamiltonians and classical
spin.  Some of the most illuminating examples are taken from the field
of simple hamiltonian mechanics.  Complicated non-local non-polynomial
hamiltonians can be considered~\cite{SchWes92,Wes99} as the
free-particle hamiltonian in appropriately chosen geometries. An
important dynamical variable which can also be considered as part of
the space-time geometry is classical spin; a relativistic spinning
particle can be described~\cite{Mad89c} as an ordinary particle
in a noncommutative geometry.  

We shall be mainly concerned with a further example of this sort,
involving an external field $B$ which can be absorbed into an
appropriate redefinition of the commutation relations of a
noncommutative geometry~\cite{Sch99}.  When considered as part of the
geometry the field $B$ changes the structure of the gauge group,
indirectly because of the way the commutation relations of the algebra
depend on it. A Yang-Mills potential $A$ has one gauge group in the
presence of a $B$ field considered as external field and its
noncommutative counterpart $\h{A}$ has another. Since the physics
cannot depend on the interpretation of the field there must be a
well-defined map $\h{A} = \h{A}(A,B)$ which reduces to the identity
when $B=0$. In the third section we shall interpret this map as a map
between covariant derivatives. We also mention the Kaluza-Klein
interpretation. The set of noncommutative structures over space-time
is in many aspects similar to a Kaluza-Klein extension. This is
particularly clear when the noncommutativity is due to a matrix
algebra~\cite{Mad89c}. The $B$ field acts then as a set of extra
coordinates which parametrize the extra dimensions. This is implicit
in earlier work~\cite{Sny47a,Mad89c} where the role of the $B$ field is
played by the spin. In fact by simply counting indices one can
conclude that extra variables are necessary. If an algebra has 4
generators then the set of all commutators has 6 elements. The
smallest algebra one can consider is the associative algebra of
dimension $10=4+6$ which is a representation of the Lie algebra of the
de~Sitter group. In the last section we present a finite model which
illuminates some of the aspects of the map. In the Appendix we recall
some basic facts about the particular version of noncommutative
geometry which will be used.  We shall set a tilde on a quantity when
it is necessary to distinguish the commutative limit.  Words in quotes
are ill-defined.

\initiate
\section{Paleoparadigmata}   

A free particle in motion in a curved space-time can be considered as
a particle in a flat space-time moving under the influence of an
external field. There is an analogous example in noncommutative
geometry.  Consider an interaction hamiltonian $H = H_0 + V$ on the
real line $\b{R}$ with time added or not. Then for appropriate $V$
these hamiltonians are equivalent~\cite{SchWes92,Wes99} to free
hamiltonians acting on often exotic noncommutative structures. Such
phenomena exist also in field theory.  There have been in the recent
literature several models which can be either considered as unified
field theories on flat space-time or as abelian gauge theory on an
appropriate noncommutative geometry.  We mention these models first as
examples of the phenomenon which we wish to investigate here because
they can also be interpreted from another closely related point of
view, that of dimensionally reduced Kaluza-Klein theories. There is a
version of this theory which involves a matrix geometry in the hidden
dimensions and so an abelian-gauge theory in the noncommutative
geometry appears as an $U_n$ gauge theory including the associated
Higgs-Kibble scalars, when regarded traditionally as an external field
problem in a plain, flat geometry. Two simple examples can be given to
illustrate how the abelian-gauge action over a noncommutative geometry
contains supplementary fields when reinterpreted in terms of ordinary
geometry.  These examples involve noncommutative extensions of the
algebra of functions on space-time.  The extra modes are hidden in the
extra structure. For simplicity of presentation we shall replace
space-time by a point and consider only the extra noncommutative
geometry.

As a first example~\cite{ConLot90} write $\b{C}^2 = \b{C}^1 \oplus \b{C}^1$ 
and decompose accordingly the algebra of $2 \times 2$ matrices
$M_2 = M_2^+ \oplus M_2^-$ into diagonal and off-diagonal parts. The
commutative algebra $M_2^+$ is the algebra of functions on 2 points.
Introduce a graded derivation $ d \alpha$ of $\alpha \in M_2$ by
$$
d \alpha = - [\eta \k \alpha], \qquad \eta \in M_2^-.
$$
The bracket is graded and $\eta$ is anti-hermitian. We find that 
$d\eta = -2\eta^2$ and that $ d^2 \alpha = [ \eta^2\k \alpha ]$. If we
choose $\eta$ such that $\eta^2 = - 1$ then $d^2 = 0$.  Then
$\Omega^*_\eta = M_2$ is a differential calculus over $M^+_2$. Notice that 
\begin{equation}
d \eta + \eta ^2 = 1.                                  \label{d-eta}
\end{equation}
Choose $\psi \in M^+_2$. A covariant derivative is given by 
\begin{equation}
D_{(0)}\psi = - \eta \psi.                              \label{sec:1}
\end{equation}
We recall that a covariant derivative must satisfy a left-Leibniz
rule. Because of the definition of $d$ one sees that this is indeed
the case:
$$
D_{(0)} (f\psi) = - \eta f\psi = df \psi - f \eta \psi.
$$
The most general $D$ is necessarily of the form 
$$
D \psi = - \eta \psi - \psi \phi
$$
where $\phi$ defines a left-module morphism of $M_2^+$. If one introduce
the map
$$
d\psi = - [\eta\k \psi]
$$
one can write $ D\psi = d\psi - \psi \omega $ in terms of a 
`connection form' $\omega = \eta + \phi$ which transforms as
$$
\omega^\prime = g^{-1} \omega g + g^{-1} dg, \quad 
g  \in U_1 \times U_1.
$$
Since in particular $ \eta^\prime = \eta$ one finds that
\begin{equation}
\phi^\prime = g^{-1} \phi g.                           \label{sec:3}
\end{equation}
The curvature is 
$$
\Omega = d\omega + \omega^2 = 1 + \phi^2  = 1 - \vert \phi \vert^2
$$
and the analogue of the abelian-gauge action is given by
$$
V(\phi) = \frac 14 \tr (1 - \vert \phi \vert^2)^2.
$$
We emphasize the fact that it is abelian-gauge theory; the geometry has
changed not the theory being studied. Because of the exotic geometry
however the result looks more like abelian Higgs theory.

As another example~\cite{DubKerMad89a,Mad89c} we consider the algebra
$M_n$ of $n\times n$ complex matrices with an anti-hermitian basis
$\lambda_a$ of $SU_n$ and define the frame
$$
\theta^a = \lambda_b \lambda^a d \lambda^b.
$$
The structure of the algebra $\Omega^*(M_n)$ is given by the relations
$\theta^a \theta^b = - \theta^b \theta^a$. These relations can be rewritten
in the form~(\ref{struc}) in the special case~(\ref{P0}). It is
easily seen that
$$
d \theta^a = - \frac 12 C^a{}_{bc} \theta^b \theta^c
$$
from which follows that
\begin{equation}
d\theta + \theta^2 = 0,\qquad \theta = - \theta^a\lambda_a.  \label{3.1.14}
\end{equation}
A special covariant derivative is given by
$$
D_{(0)}\psi = - \theta \psi
$$
and the most general one is of the form 
$$
D \psi = - \theta \psi - \psi \phi.
$$
If one introduce the map
\begin{equation}
d\psi = - [\theta\k \psi]                                     \label{3.1.15}
\end{equation}
one can write again $D\psi = d\psi - \psi\omega$ in terms of a 
`connection form' $\omega = \theta + \phi$ which transforms as
$$
\omega^\prime = g^{-1} \omega g + g^{-1} dg, \quad g \in U_n.
$$
Since in particular $\theta^\prime = \theta$ one finds again~(\ref{sec:3}).
The curvature is
$$
\Omega = d\omega + \omega^2 = 
\frac 12 \Omega_{ab} \theta^a \theta^b
$$
where
$$
\Omega_{ab} = [\phi_a\k\phi_b] - C^c{}_{ab}\,\phi_c.
$$
The $C^c{}_{ab}$ is a sort of `Christoffel symbol'; the algebra $M_n$
with the present differential calculus is `curved' as a geometry. The
analogue of the electromagnetic action is
$$
V(\phi) = \frac 14 \tr (\Omega_{ab} \Omega^{ab}).
$$
Again, as above, this action describes `abelian-gauge' theory on a
noncommutative `space'. By radically changing the `space' we have
radically changed the aspect of a well-known theory. 

We have presented these two examples in some detail since they
illustrate well the definition of a covariant derivative. In both
cases the module is a bimodule over the algebra.  The covariant
derivative however uses only the right-module structure and satisfies
a right-Leibniz rule. The left-module structure is reserved for the
action of the gauge group which we identify as a subset of the
algebra. We shall encounter similar calculations in the next section.

As examples of noncommutative extensions of space-time we shall choose
algebras which are deformations of the algebra of smooth functions on
Minkowski space. Let $\t{x}^{\mu}$ be cartesian coordinates. As has been
done previously~\cite{Sny47a,Mad89c,DopFreRob95} we replace $\t{x}^{\mu}$
by four hermitian generators $x^{\mu}$, elements of an abstract
$*$-algebra $\c{A}$ which do not commute:
\begin{equation}
[x^{\mu}\k x^{\nu}] = i\kbar J^{\mu\nu},   \qquad  
x^{\mu *} = x^{\mu}.                                          \label{sec:2}
\end{equation}
The parameter $\kbar$ is so chosen so that $J^{\mu\nu}$ has no
dimensions.  We shall set $\kbar = 1$ by a choice of units.
A natural Ansatz which respects all reflection symmetries would be
\begin{equation}
x^{\mu} = \t{x}^{\mu} + \kappa J^{\mu}, \qquad
J^{\mu} = \bar z \gamma^{\mu} z.                               \label{8.1.4}
\end{equation}
We shall impose on $z$ the following commutation relations:
\begin{equation}
[ z\k z] = 0, \quad [z\k \bar z ] = 1, \quad [\bar z\k \bar z ] = 0. 
                                                               \label{8.1.5}
\end{equation}
The unit on the right-hand side of these equations is the tensor
product of the unit in the Clifford algebra and the unit in the
operator algebra.  Written out in terms of components of the Dirac
spinors Equations~(\ref{8.1.5}) become
$$
[ z^\alpha\k z^\beta] = 0, \quad 
[z^\alpha\k \bar z_\beta ] = \delta^\alpha_\beta, \quad 
[\bar z_\alpha\k \bar z_\beta ] = 0. 
$$
If we introduce
$$
S^{\mu\nu} = \bar z \sigma^{\mu\nu} z, \qquad 
\sigma^{\mu\nu} = {i\over 2} [\gamma^{\mu}\k \gamma^{\nu}]     
$$
then from the commutation relations (\ref{8.1.5}) follow the commutation
relations (\ref{sec:2}) for the generators with 
$$
S^{\mu\nu} = - J^{\mu\nu},\qquad \kbar  = 2 \kappa^2.     
$$ 
We can consider the Dirac spinor as an element of the quantized
version of an algebra of functions over the classical phase space 
$(z\k \bar z)$ with Poisson bracket $\{ z\k \bar z \} = i$. There are
therefore two distinct quantization procedures, the
ordinary one involving $\hbar$ and this new one. As a mathematical
simplification we shall `dequantize' $z$ and consider the classical
phase space $(z\k \bar z)$. Introduce $C^{\lambda}$ by
$$
C^\lambda = {i\over 2} 
\big(( \bar z \gamma^\lambda)_\alpha \bar \p^\alpha 
  - (\gamma^\lambda z )^\alpha \p_\alpha \big),  \quad 
\p_\alpha = \p /\p z^\alpha                              
$$
and consider the condition
\begin{equation}
\p_{\lambda} C^{\lambda} f = 0.                        \label{8.1.10}
\end{equation}
This is of second order in all the derivatives but of first order in
$\p_{\lambda}$. So it resembles a constraint. If $f$ depends
only on the quantity $x^{\lambda}$ defined in (\ref{8.1.4}) then
(\ref{8.1.10}) is identically satisfied. However, the converse is not
true. To the $(\t{x}^{\mu}, z, \bar z)$ we add $p_{\lambda}$ to form a phase
space. We extend the bracket by requiring that
$(p_{\lambda}, \t{x}^{\mu})$ Poisson-bracket-commute with $(z, \bar z)$. It
is not this full phase space which interests us but rather the reduced
phase space given by the $(p_{\lambda}, x^{\mu}, z, \bar z)$ which
satisfy the constraints (\ref{8.1.10}). This reduced phase space
describes the motion of a spinning particle. Define $S^{\lambda}$ by
$$
S^{\lambda} = \bar z \gamma^{\lambda} \gamma^5 z.           
$$
Then the constraints (\ref{8.1.10}) are equivalent to the conditions
$$
\begin{array}{lll}
p^2 - \mu^2 = 0, &p_{\mu} S^{\mu} = 0, &\bar z \gamma^5 z = 0, \\[4pt]
\mu J^{\lambda} = \bar z z p^{\lambda},
&\mu S^{\mu\nu} = \epsilon^{\mu\nu\rho\sigma} p_{\rho} S_{\sigma}.
\end{array}
$$
The parameter $\mu$ is a mass parameter. 

Models can be constructed using the tensor product, for example using
the algebras introduced in (\ref{matrices}). We shall need to slightly
change our notation since the situation we consider here is very
similar to the situation of the next section in which $\c{A}$ and
$\h{\c}{A}$ describe noncommutative versions of flat space-time or of
a brane and the matrix factor is a modified Kaluza-Klein
extension~\cite{Mad89c}.  Let $\c{A}$ and $\c{B}$ be two algebras with
differential calculi $\Omega^*(\c{A})$ and $\Omega^*(\c{B})$. Then
there is a natural differential calculus over the $\c{A} \otimes
\c{B}$ given by
\begin{equation}
\Omega^*(\c{A} \otimes \c{B}) = 
\Omega^*(\c{A}) \otimes \Omega^*(\c{B}).                   \label{3.3.1}
\end{equation}
If $\alpha \in \Omega^* (\c{A})$, $\beta \in \Omega^p (\c{B})$,
$\gamma \in \Omega^q (\c{A})$ and $\delta \in \Omega^* (\c{B})$ then
the product in $\Omega^*(\c{A}) \otimes \Omega^*(\c{B})$ is given by
\begin{equation}
(\alpha \otimes \beta)(\gamma \otimes \delta) 
= (-1)^{pq} \alpha\gamma \otimes \beta\delta.             \label{3.3.2}
\end{equation}
Equation~(\ref{3.3.1}) does not define the only choice of differential
calculus over the product algebra. Consider the module of 1-forms
$$
\Omega^1(\c{A} \otimes \c{B}) = \c{A} \otimes \Omega^1(\c{B})
\oplus \Omega^1(\c{A}) \otimes \c{B}.
$$
It can be used to construct another differential calculus
$\Omega^*(\c{A} \otimes \c{B})$ over the tensor product of the two
algebras which is in a sense the largest which is consistent with the
module structure. This extension is in general larger than the tensor
product. If $\theta^\alpha$ is a frame for $\Omega^1(\c{A})$ and
$\theta^a$ is a frame for $\Omega^1(\c{B})$ then
$$
(\theta^\alpha, \theta^a) = (\theta^\alpha \otimes 1, 1 \otimes
\theta^a)
$$
is a frame for $\Omega^1(\c{A} \otimes \c{B})$. The commutation
relations for each factor can be extended to the entire frame by the
rule (\ref{3.3.2}). In this case both constructions yield the same
algebra of forms. We are interested in the case with $\c{B} =
M_n$. Then if we define
$$
\Omega^1_h = \Omega^1(\c{A}) \otimes M_n, \qquad \Omega^1_v = \c{A}
\otimes \Omega^1(M_n),
$$
we can write $\Omega^1(\c{A}\otimes M_n)$ as a direct sum:
$$
\Omega^1(\c{A}) =  \Omega^1_h \oplus \Omega^1_v.             
$$
The differential $df$ of an element $f$ of $\c{A}$ is given by
$$
df = d_hf + d_vf.                                            
$$
We have written it as the sum of two terms, the horizontal and
vertical parts, using notation from Kaluza-Klein theory.  The algebra
$\Omega^*(\c{A})$ of differential forms is given in terms of the
differential forms of each factor by the formula:
\begin{equation}
\Omega^p(\c{A}\otimes M_n) = 
\bigoplus_{i+j = p} \Omega^i(\c{A})
\otimes \Omega^j(M_n).                                        \label{p-cal}
\end{equation}
Consider two elements $f,\,g \in \c{A}\otimes M_n$.  Let $x^\mu$ be
the generators of $\c{A}$ and use the Gell-Mann matrices $\lambda_a$
as a basis of $M_n$, as described in the Appendix. If we expand 
$f = f^0 + f^a \lambda_a$ and $g = g^0 + g^a \lambda_a$ then we find that
the commutator is given by
$$
[f\k g] = \frac 12 [f^a \k g^b] F^c{}_{ab} \lambda_c + \frac 12
[f^a \k g^b] D^c{}_{ab} \lambda_c + \frac 1n [f^a \k g^b] g_{ab} +
([f^0\k g^a] - [g^0\k f^a]) \lambda_a.
$$
As a set of generators for the algebra we can choose the tensor
products $x^\mu \otimes 1$ and $1\otimes \lambda^a$. These would
correspond respectively in Kaluza-Klein theory to the space-time
coordinates and the internal coordinates. The commutation relations
for the two sets follow immediately from (\ref{g-c-r}), with an
appropriate change of notation.

An interesting example can be found~\cite{JurSch00} using group
manifolds. A group manifold $M_G$ can be embedded as a submanifold of
its Lie algebra considered as an euclidean space. Let $x^i$ be the
coordinates of this space and consider the Poisson bracket defined by
the Lie bracket. The procedure of star quantization will yield once
again the Lie bracket. If the group is compact all irreducible
representations will be of finite dimension; there are an infinite
number indexed by Casimir operators $c_i$, each with a well-defined
dimension $d_i$. If we set $\c{A} = \c{C}(M_G)$ then we can write
$$
\h{\c}{A} = \bigoplus_{c_i} M_{d_i}.
$$
This situation generalizes to arbitrary K\"ahler
manifolds~\cite{Sch99}. We are especially interested in situations which
at least in some formal sense we can identify
\begin{equation}
\underline{\h{g}} \sim \bigoplus_{i} \underline{su}_{n_i}.
\label{sum}
\end{equation}
If the algebra $\h{\c}{A}$ contains a matrix algebra $M_n$ then one
can consider $\underline{su}_{n}$ as a subalgebra of
$\underline{\h{g}}$.

\initiate
\section{Noncommutativity versus Field Theory}

Consider again the formal algebra $\c{A}$ of the previous section
defined less precisely in terms of commutation relations of the
form~(\ref{sec:2}) but with the right-hand side a non-specified set of
elements of the algebra. Consider also a second algebra $\h{\c}{A}$
which has the same number of generators $\h{x}^i$ but in general a
different set of elements $\h{J}^{ij}$ on the right-hand side of the
commutation relations. We shall suppose that both of these algebras
can be represented as subalgebras of the algebra of differential
operators $\t{\c}{A}$ on some space of smooth functions. In the
Appendix such a representation is given explicitly in a special but
important case. We designate the product in $\c{A}$ by $*$ and in
$\h{\c}{A}$ by $\hat *$.

We assume that there is an algebra homomorphism
\begin{equation}
\h{\c}{A} \buildrel \rho \over \longrightarrow \c{A}      \label{sec:5}
\end{equation}
of $\h{\c}{A}$ onto $\c{A}$ which can be formally defined by the action
$$
x^i = \rho (\h{x}^i) = \Lambda^i(\h{x}^j)
$$
on the generators. By assumption then
\begin{equation}
\rho(\h{x}^i \hwm \h{x}^j) = x^i \wm x^j = 
\rho(\h{x}^i) \wm \rho(\h{x}^j).                      \label{cop}
\end{equation}
The kernel of $\rho$ is a 2-sided ideal so $\h{\c}{A}$ cannot in any
sense of the word be `simple'. If $\h{\c}{A}$ is commutative then so
obviously is $\c{A}$; if on the other hand $\c{A}$ is commutative then
the kernel of $\rho$ contains necessarily the ideal generated by the
commutators. If $\h{J}$ is non-degenerate then this can again by
identified with $\c{A}$ and so $\rho = 0$.  In the special case with
$J$ and $\h{J}$ constant non-degenerate matrices we can choose
$F^i(\h{x}^j) = F^i_j \h{x}^j$ a linear transformation.
We have then
$$
x^i \wm x^i = F^i_k F^j_l \h{x}^k \hwm \h{x}^l, \qquad
J^{ij} = F^i_k F^j_l \h{J}^{kl}.
$$
In general the relation between the generators is much more
complicated. If we can write for example 
$F^i(\h{x}^j) = \h{x}^i - \xi^i(\h{x}^j)$ as a linear 
perturbation then
$$
i \kbar J^{ij} = [x^i \k x^j] = 
[\rho(\h{x}^j), \rho(\h{x}^j)] =
i \kbar \h{J}^{ij} - \kbar [x^{[i} \k \xi^{j]}].
$$
which we write in the form
\begin{equation}
\h{J}^{ij} = J^{ij} + \theta^{ij}, \qquad 
i \kbar \theta^{ij} = - i [x^{[i}\k \xi^{j]}].              \label{theta}
\end{equation}
If we suppose that $J^{ij}$ is constant then using the $\h{\lambda}_a$ of
the Appendix and writing $\xi^i$ as $\xi^i = i\kbar J^{ia} a_a$ we find that
$$
\theta^{ij} =  \kbar J^{ia} J^{jb} e_{[a} a_{b]}.
$$
In this case the perturbation of the commutation relations is related to
the exact form 
$$
f = da, \qquad a = a_b \theta^b, \qquad 
f = \frac 12 e_{[a} a_{b]} \theta^a \theta^b.
$$
We show in the Appendix that $d\theta^a = 0$. We refer to the
literature~\cite{Sch99} for a description of the relation between
$\theta^{ij}$ and the $B$ field.

One might be tempted to consider the $F^i(\h{x}^j)$ as a `change
of coordinates'. But the change is in the `phase space' of which
$\t{\c}{A}$ is the structure algebra and so when one looks for a
similar transformation in ordinary geometry one must imagine not only
a change of coordinates but also a shift in the position because of
the term in the definition of the generators which depends on the
momentum.  What can more properly be considered as a change of
coordinates is an automorphism of the algebra, for example the inner
automorphism
$$
\h{x}^i = \Lambda^{-1} x^i \Lambda.
$$
In this case the product is conserved. 

It is perhaps preferable to consider $\rho$ as a change of product on
one fixed vector space. We drop then the hat on the generators and
distinguish the two products by putting a hat on one of them.  In the
case of a linear perturbation Equation~(\ref{cop}) becomes
$$
\rho(x^i \hwm x^j) = x^i \wm x^j + x^i \wm \xi^j + \xi^i \wm x^j 
$$
The requirement that the new product be associative places
restrictions~\cite{Ger64} on the $\xi^i$.

In general one can consider the set $S_0$ of all products on the
vector space $\c{A}$. There is a subset $S_1 \subset S_0$ in which the
product is associative; this is the set which interests us here. Let
$\pi$ be a given product and consider the orbit $S_2 \subset S_1$ of
$\pi$ under the group of all possible maps $\rho$. This group has a
subgroup of automorphisms of $\c{A}$, which leave the product
invariant. In a formal sense $S_2$ can be identified with the quotient
of the two groups. In general $S_1$ will be a union of orbits of
different products of non-isomorphic algebras. If we assume that there
are no relations other than the commutation relations~(\ref{sec:2})
then the set $S_2$ will be parameterized by the $J^{ij}$. To pass from
stratum of $S_1$ to another would require a singular variation in $J$.
A familiar example from the theory of Lie algebras is furnished by the 
embedding $SU_n \hookrightarrow SO_{n^2-1}$. If $\{\lambda_i\}$ is a
set of generators of the Lie algebra of $SU_n$ then so is the set
$\{\h{\lambda}_i\}$ with $\h{\lambda}_i = g^{-1} \lambda_i g$ for 
$g \in SU_n$. One can write then 
$\{\h{\lambda}_i\} = \Lambda_i^j \lambda_j$ where the transformation
coefficients are complex numbers. It is the analog of those
transformations of $SO_n$ which do not respect the Lie algebra
structure which interests us here.

As a limiting case with singular $\rho$ one consider an algebra 
$\h{\c}{A}$ with a non-degenerate $\h{J}$ and an algebra $\c{A}$ with 
$J=0$. In the latter case we can
identify $x^i$ with $\t{x}^i$, the `space' coordinates of 
$\t{\c}{A}$. The `lift' by the inverse of $\rho$ is a
quantization procedure, a way of associating an operator to a
function.  One such method is the Weyl-Moyal quantization
procedure~\cite{Wey50,Moy49} which furnishes a `natural' right inverse 
for $\rho$ which lifts an element $f \in \c{A}$ to an element 
$\h{f} \in \h{\c}{A}$. This is a map between two different strata of
$S_1$.

Let $\c{H}$ be a right $\c{A}$-module and $\h{\c{H}}$ be
a right $\h{\c}{A}$-module.  We shall place a hat on an element of $\c{H}$
whenever it is necessary to distinguish the $\h{\c}{A}$-module structure.
For simplicity we shall suppose that both modules
are free over their respective algebras and so the map $\rho$ can be
extended to a map
$$
\h{\c{H}} \buildrel \rho \over \longrightarrow \c{H}
$$
between the two of them. We shall simplify even further and suppose
that the module is of rank one. It can be identified therefore with
the respective algebra and each identification is equivalent to a
choice of gauge. We choose $\psi_0 \in \c{H}$ as basis of $\c{H}$
as both $\c{A}$-module and $\h{\c}{A}$-module and we write
$\psi = \psi_0 \wm f$ and $\h{\psi} = \psi_0 \hwm f$. This defines the
map $\rho$ in terms of the products.  We shall suppose that the
potential $A$ lies in the Lie algebra $\underline{g}$ of a Lie
(pseudo)group $\c{G}$ which we shall take to be a subgroup of the
unitary elements of $\c{A}$ and likewise that $\h{A}$ lies in the Lie
algebra $\underline{\h{g}}$ of a Lie (pseudo)group $\h{\c}{G}$.  We
shall suppose that the gauge group acts on the left. The left action
of $\c{G}$ on $\c{H}$ is compatible with the algebra action from the
right. This condition is automatic in normal Yang-Mills theory where
the two actions always commute.  Since the derivative is covariant
from the left one has also
$$
D (g^{-1} \psi) = g^{-1} D\psi, \qquad g \in \c{G}.
$$
If $g \simeq 1 + h$ then one can write this in the form of a left
Leibniz rule for $h$.

In ordinary geometry the case we are considering would be called an
abelian gauge theory.  This is in fact more general since
gauge theory with unitary groups can be incorporated simply by the
replacements
\begin{equation}
M_n \otimes \c{A} \mapsto  \c{A}, \qquad
M_n \otimes \h{\c}{A} \mapsto \h{\c}{A}.                  \label{matrices}
\end{equation}
It is only important that the matrix factor be the same for both
algebras since otherwise the map $\rho$ in general would not be
interesting.  If we choose the differential calculus given by
(\ref{p-cal}) and make the replacement (\ref{matrices}) then we can
consider Equation~(\ref{SW-bis}) below to be valid also in the product
case.  The bracket must be chosen to be that of the product algebra.

We suppose finally that there is a differential calculus
$\Omega^*(\c{A})$ over $\c{A}$ and a differential calculus
$\h{\Omega}^*(\h{\c}{A})$ over $\h{\c}{A}$ and that the map $\rho$ can
be extended to an algebra morphism
$$
\h{\Omega}^*(\h{\c}{A}) 
\buildrel \rho \over \longrightarrow \Omega^*(\c{A})
$$
of the latter onto the former.  As important special cases we
mention the calculi whose modules of 1-forms are free with a special
basis (frame) $\theta^a$ and $\h{\theta}^a$ as given in the Appendix.
We have then the identifications
$$
\Omega^1(\c{A}) = \bigoplus_1^d \c{A}, \qquad
\h{\Omega}^1(\h{\c}{A}) = \bigoplus_1^d \h{\c}{A}.
$$
The integer $d$ here is the `dimension' and must be the same in
both cases. The extension of $\rho$ can be defined by setting
\begin{equation}
\rho (d \h{f}) = d\rho (\h{f}).                             \label{rhost}
\end{equation}
This is a natural extension but it is not necessarily compatible with
the identification of a form with its components. The image of a free
module is not necessarily free.

Let $D$ and $\h{D}$ be covariant derivatives defined on respectively
$\c{H}$ and $\h{\c{H}}$.  We introduce the gauge potentials as usual
by the conditions
$$
D \psi_0 = \psi_0 \wm A, \qquad \h{D} \psi_0 = \psi_0 \hwm \h{A}.
$$
These define $D$ and $\h{D}$ on all of $\c{H}$ either by the
Leibniz rule or by the gauge covariance. If $f \simeq 1 + h$ then to
first order in $h$ we can write
$$
D\psi = \psi \wm (A + Dh), \qquad 
\h{D}\psi = \psi \hwm (\h{A} + \h{D} h).
$$
We have here introduced the covariant derivatives
$$
Dh = dh + [A\k h], \qquad \h{D} h = \h{d}h + [\h{A}\hk h]
$$
of an element $h \in \c{A}$ ($\h{\c{A}}$), with
$$
[A\k h] = A\wm h - h\wm A, \qquad
[\h{A}\hk h] = \h{A} \hwm h - h \hwm \h{A}.
$$
Conversely, given $A$ and $\h{A}$ one can construct a
map~\cite{SeiWit99}
$$
\mbox{SW}: \; D \longrightarrow \h{D}
$$
between the two derivatives by assuring that the two Leibniz rules
are satisfied.  The map SW becomes then an equation because of
integrability conditions; it must be well-defined on all of $\c{H}$.

If $\rho$ is an automorphism then $\h{D} - D$ is a (right) module
morphism. One can neglect the distinction between the two products and
write
\begin{equation}
\h{D} h = D h + [\Gamma\k h]                           \label{Gamma0}
\end{equation}
with $\Gamma = \h{A} - A$. 
If we define the variation 
\begin{equation}
\delta_h \Gamma = \h{D} h - D h
\end{equation}
of $\Gamma$ under multiplication by $f \simeq 1 + h$, we see that it 
is given by
\begin{equation}
\delta_h \Gamma = [\Gamma\k h].                         \label{Gamma}
\end{equation}
This is the well-known formula which expresses the gauge covariance of
the difference between two connections. The map SW is a generalization
of this formula to situations where the two connections in question
are with respect to two different gauge groups.

In general, if $\rho$ is not an automorphism, then
Equation~(\ref{Gamma}) will have no solution and we cannot define
$\Gamma$ as we have done. Since $\rho$ is surjective we can introduce
a function $\gamma(h)$ with values in $\c{A}$ such that 
$$
\psi_0 \hwm (1 + h) = \psi_0 \wm (1 + h)(1 + \gamma).
$$
This implies that $\psi_0 \hwm dh = \psi_0 \wm d(h + \gamma)$ and
therefore that
$$
\h{D} \psi = \psi \wm \h{D} (h + \gamma[h]).
$$
Using the definition of $\delta_h \Gamma$ given above this can be
written as
\begin{equation}
\delta_h \Gamma = D\gamma + \h{D} h - D h = 
D\gamma  + [\Gamma \k h] + [\h{A} \hk h] - [\h{A} \k h].         \label{SW}
\end{equation}
If $\rho$ is not an automorphism then to compensate for the difference
between $\rho$ and an automorphism we have introduced an element
$\gamma \in \underline{g}$.  This is equivalent to an interpretation
of the modification of the product by a change of gauge. We have in
fact identified the gauge group as the unitary elements of the
algebra. When we change the structure of the algebra this entails
necessarily a change in the structure of the gauge group and hence of
the Lie algebra. In certain cases the change involves a finite number
of parameters in the commutation relations. As an example of this one
can consider (\ref{theta}) with the $\theta^{ij}$ real numbers. A
gauge transformation which depends on these extra parameters is
equivalent to a local gauge transformation in a Kaluza-Klein extension
of the theory with the $\theta^{ij}$ as the local coordinates of the
extra dimensions. The variation described in Equation~(\ref{SW}) is
however for fixed `Kaluza-Klein' parameters and gives only the
variation of $\Gamma$ under change of gauge. Having found the
solution explicitly in terms of the extra parameters one could
calculate also their variation. 

Both $D$ and $\h{D}$ can be extended to the entire differential
calculus; in general however there is no extension of SW.  In the
special cases we are considering here both of the differential calculi
can be written in the form
$$
\Omega^*(\c{A}) = \c{A} \otimes \textstyle{\bigwedge}^*
$$
where the second factor is the deformed exterior algebra over the
vector space spanned by the frame. If
$$
\textstyle{\bigwedge}^* = \h{{\textstyle{\bigwedge}}}{}^*
$$
then both $\rho$ and SW can be extended to the exterior algebra.
We can write
$$
D \psi = \theta^a D_a \psi, \qquad 
\h{D} \psi = \h{\theta}^a \h{D}_a \psi.
$$
We shall restrict our attention here to the important special case
with the projector $P^{ab}{}_{cd}$, defined in the Appendix, given by
the expression~(\ref{P0}). We have then
\begin{equation}
[\lambda_a \k \lambda_b] = 
\lambda_c F^c{}_{ab} + K_{ab}, \qquad
[\lambda_a \hk \lambda_b] = 
\lambda_c \h{F}^c{}_{ab} + \h{K}_{ab}.                  \label{g-c-r}
\end{equation}
It follows from~(\ref{struc}) that the product structure of the frame
is the same with or without hat. One finds from~(\ref{W-M}), to lowest
order, the expression
$$
\h{e}_a f = e_a f + 
i\kbar \theta^{bc} [\lambda_b \hk \lambda_a] \hwm \h{e}_c f
$$
for the `partial derivatives'. As seen by comparing~(\ref{g-c-r})
with~(\ref{K-J}), this is an identity.  The frame is gauge invariant:
$\delta_h \theta^a = 0$.  Because of the special properties of the
frame Equation~(\ref{SW}) can be written using components as
\begin{equation}
\delta_h\Gamma_a = D_a \gamma + [\Gamma_a \k h] + 
\frac 12 \theta^{bc} [e_b A_a \k e_c h] + 
 + o(\kbar^2).                                               \label{SW-bis}
\end{equation}
The solution is difficult to find in general but if the deformation
parameter $\kbar$ which defines the algebra $\h{\c}{A}$ in terms of
$\c{A}$ is small a formal Taylor-series expansion can be
given~\cite{SeiWit99}.  In the limit then $J^{ab} \to 0$
Equation~(\ref{SW-bis}) can be written using only ordinary derivatives
as
\begin{equation}
\delta_h\Gamma^a = \theta^{aj} D_j \gamma + [\Gamma^a \k h] - 
\frac 12 \theta^{kl} [\p_k h \k \p_l (\theta^{aj} a_j)], 
\qquad \Gamma^a = \theta^{ab}\Gamma_b.                          \label{Wess}
\end{equation}
To emphasize the special status of this case we have written the
potential using a lower-case letter: $A_i \mapsto a_i$. 

In principle the preceding must be generalized to the case where the
covariant derivative includes a gravitational contribution. We have
changed the structure of the algebra without changing that of the
differential calculus and this is not always possible.  With the
formalism we have used, based on the existence of a frame we have
essentially assumed that the differential calculus is not gauge
dependent. In general this will not be true since the gauge group
depends on the structure of the algebra and the differential calculus
depends on the latter.  The pair $(\gamma,\Gamma)$ of external fields
depends through Equation~(\ref{SW}) on the Poisson structure $\theta$
which in turn can be identified with the $B$ field.  One can say then
that the map SW is another example of the equivalence between the
point of view which considers geometry as an essential given aspect of
space-time and the point of view which considers geometry as a
convenient description of an external field on a conventional
space-time. In other words we are lead to interpret SW as a
correspondence between on the one hand some physical situation with
external fields and on the other the same physics but with the extra
variables considered as an intrinsic part of a noncommutative
geometry.

\initiate
\section{Neoparadigma} 

In this section we shall consider an example of the map SW constructed
using the first two examples of Section~2. This will consist in a
contraction of the second model onto the first~\cite{MadMouSit97}.
The algebras are respectively 
$$
\h{\c{A}} = M_2, \qquad \c{A} = M^+_2.
$$
One can think of the limit as the classical limit of a quantum spin
or as a contraction of a gauge group.  The `local' gauge group of the
algebra $M_2$ is the group $U_2$ and that of $M_1 \times M_1$ is 
$U_1 \times U_1$.  Associated to the latter are two gauge potentials,
the photon $\gamma$ and a massive neutral vector boson $Z_0$; the
former has also a massive charged $W$.  The contraction can be
implemented by letting the $W$ mass tend to infinity.  The role of the
$B$-field is played by the charged $W$-boson. In this example there is
no obvious interpretation of the commutation relations of $\h{\c{A}}$
in terms of a $B$-field, unless it be the fact that the $W$-boson
takes its values in the complement of $U_1 \times U_1$ in $U_2$.  The
passage from $\c{A}$ to $\h{\c{A}}$ is here an example of a map
between algebras which is not a deformation quantization.

We introduce $\rho_\epsilon$ by the action
$$
\rho_\epsilon (\h{\lambda^1}) = \epsilon \lambda^1, \qquad
\rho_\epsilon (\h{\lambda^2}) = \epsilon \lambda^2, \qquad
\rho_\epsilon (\h{\lambda^3}) = \lambda^3
$$
on the Pauli matrices. Therefore the structure constants rescale as
$$
C^1{}_{23} = \h{C}^1{}_{23}, \qquad 
C^2{}_{31} = \h{C}^2{}_{31}, \qquad
C^3{}_{12} = \epsilon^{-2} \h{C}^3{}_{12}
$$
and the metric as $g^{ab} = \mbox{diag}(\epsilon^2, \epsilon^2, 1)$.
For all $\epsilon > 0$ this is a redefinition of the product of
$M_2$ such that $\rho_\epsilon$ is an isomorphism and for 
$\epsilon = 0$ it is a singular contraction.  We define $\rho_0$ to be
the singular limit as $\epsilon \to 0$. If we decompose 
$\h{f} = \h{f}^+ + \h{f}^-$ then we have 
$\rho_\epsilon(\h{f}) = f^+ + \epsilon f^-$ and
$$
\rho_\epsilon(\h{f} \hwm \h{g}) = f^+ \wm g^+ + o(\epsilon).
$$
It follows that the image of $\rho_0$ contains nilpotent elements. This
accounts for the difference in the dimensions of $\h{\c{A}}$ and $\c{A}$.
Except for a rescaling the frame remains invariant under the
contraction and the extension~(\ref{rhost}) is given simply by
$$
\theta^1 = \epsilon \h{\theta}^1, \qquad
\theta^2 = \epsilon \h{\theta}^2, \qquad
\theta^3 = \h{\theta}^3.
$$
The differential remains invariant: 
$$
\rho_\epsilon (\h{d}\h{f}) = d \rho_\epsilon(\h{f}).
$$

We choose $\psi_0 = 1$, the unit matrix of $M_2$ and we set 
$\h{D} \cdot 1 = \h{A} = \h{A}_a \h{\theta}^a$. The image $\h{A}$
under $\rho_\epsilon$ must be of the form 
$\rho_\epsilon(\h{A}) = A_3(\lambda^3) \theta^3 + o(\epsilon)$. The
remaining two modes become infinitely heavy in the limit and decouple.
With the identifications it follows that near the identity matrix we
can write $\h{h} = h + \gamma$.  We can therefore write
$$
\h{D} \h{h} = d(h + \gamma) + [\h{A} \k \gamma] + [\h{A} \hk h], 
\qquad Dh = dh
$$
and (\ref{SW}) becomes the equation
$$
\delta_h \Gamma = d\gamma + [\h{A} \k \gamma] + [\h{A} \hk h].
$$
Since $h$ defines a gauge transformation of $A$ it must be of the
form $h = h_3\lambda^3$. If therefore $\h{A} = \h{A}(\h{\lambda^3})$
then a solution is given by $\gamma = 0$, $\Gamma = 0$.  One can
consistently choose $\h{A} = A$. If on the other hand
$$
\h{A} = \h{A}_3(\h{\lambda}^1,\h{\lambda}^2) \h{\theta^3},
$$ 
for example, then the equation becomes the equation
\begin{equation}
\delta_h \Gamma_3 = e_3 \gamma + [A_3, \gamma] + 
[\h{A}_3 \hk h]                                          \label{SW-eg}
\end{equation}
for the third component. The source term $[\h{A}_3 \hk h]$ now is not
equal to zero and the external fields, the difference between the
potentials $\Gamma_3$ as well as the `scalar' $\gamma$, cannot vanish.
We are free to interpret them as components in a noncommutative
geometry or as external fields in a commutative (albeit discrete) one.

\initiate
\section*{Acknowledgments} 

The authors would like to thank Paolo Aschieri, Gaetano Fiore and
Harold Steinacker for enlightening conversations.

\initiate
\section{Appendix} 

Let $\c{A}$ be a noncommutative algebra with a differential calculus
$\Omega^*(\c{A})$.  A large class of differential calculi, but not
all, are such that the module $\Omega^1(\c{A})$ is free as a left or
right $\c{A}$-module and has a special frame $\theta^a$ with
\begin{equation}
[f\k \theta^a] = 0, \qquad 1 \leq a \leq n                 \label{frame}
\end{equation}
which is dual to a set of derivations $e_a = \ad \lambda_a$:
\begin{equation}
df = e_a f \theta^a = [\lambda_a\k f] \theta^a = 
- [\theta\k f], \qquad \theta = -\lambda_a \theta^a.         \label{dual}
\end{equation}
The set of $\theta^a$ is the noncommutative equivalent of a Cartan
moving frame and in ordinary geometry the derivations $e_a$ would be
called Pfaffian derivatives. The `Dirac operator' $\theta$ generates
$\Omega^1(\c{A})$ as a bimodule; it is not a free bimodule.  The
$\lambda_a$ must satisfy the consistency condition~\cite{DimMad96}
\begin{equation}
2 \lambda_c \lambda_d P^{cd}{}_{ab} - 
\lambda_c F^c{}_{ab} - K_{ab} = 0.                           \label{fund-eq}
\end{equation}
It has been shown recently~\cite{MadSchSchWes00a} that this can be
interpreted as a vanishing-curvature condition.

The $P^{cd}{}_{ab}$ define the product in the algebra of forms:
\begin{equation}
\theta^a \theta^b = P^{ab}{}_{cd} \theta^c \theta^d.            \label{struc}
\end{equation}
The $F^c{}_{ab}$ are related to the 2-form $d\theta^a$ through the
structure equations:
\begin{equation}
d\theta^a = - {1\over 2}C^a{}_{bc} \theta^b \theta^c, \qquad
C^a{}_{bc} = F^a{}_{bc} - 2 \lambda_e P^{(ae)}{}_{bc}.           \label{s-e}
\end{equation}
The $K_{ab}$ are related to the curvature of $\theta$:
$$
d\theta + \theta^2 = {1\over 2} K_{ab} \theta^a \theta^b.
$$
All the coefficients lie in the center $\c{Z}(\c{A})$ of the
algebra.  With no restriction of generality we can impose the
conditions
\begin{equation}
F^e{}_{cd} = P^{ab}{}_{cd} F^e{}_{ab}, \qquad K_{cd} = 
P^{ab}{}_{cd} K_{ab}.                                           \label{F-P-K}
\end{equation}
Define
$$
C^{ab}{}_{cd} = \delta^a_c \delta^b_d - 2 P^{ab}{}_{cd}.       
$$
Then from the fact that $P^{cd}{}_{ab}$ is a projector we find that
$C^{ab}{}_{cd}C^{cd}{}_{ef} = \delta^a_e \delta^b_f$.  We can write
then the first term of Equation~(\ref{fund-eq}),
$$
2 \lambda_d \lambda_e P^{de}{}_{bc} = \lambda_b \lambda_c -
\lambda_d \lambda_e C^{de}{}_{bc} \equiv
[\lambda_b\k \lambda_c]_C,                                       
$$
as a sort of deformed bracket and Equation~(\ref{fund-eq}) can be
rewritten in the form
\begin{equation}
[\lambda_b\k \lambda_c]_C = \lambda_a F^a{}_{bc} + K_{bc}.
\label{consis}
\end{equation}
If $P^{ab}{}_{cd}$ is given by
\begin{equation}
P^{ab}{}_{cd} = {1 \over 2} (\delta^a_c \delta^b_d - \delta^a_d
\delta^b_c) \label{P0}
\end{equation}
then we have
$$
C^{ab}{}_{cd} = \delta^b_c \delta^a_d.                          
$$
Equation~(\ref{consis}) defines a `twisted' Lie algebra with a
central extension and the $F^a{}_{bc}$ must satisfy a set of modified
Jacobi identities.  From (\ref{consis}) one derives immediately the
relations
\begin{equation}
[e_a\k e_b]_C = C^c{}_{ab} e_c.  \label{Lie}
\end{equation}
between the first and second derivatives. When $P^{ab}{}_{cd}$ is of
the form (\ref{P0}) the derivations form a Lie algebra.

As an example we recall the case of the matrix algebra $M_n$.  Let
$\lambda_a$, for $1 \leq a \leq n^2-1$ be an anti-hermitian frame of
the Lie algebra of the special unitary group $SU_n$. The product
$\lambda_a \lambda_b$ can be written in the form
\begin{equation}
\lambda_a \lambda_b = {1\over 2} F^c{}_{ab} \lambda_c + {1\over 2}
D^c{}_{ab} \lambda_c - {1 \over n} g_{ab}.  \label{3.1.1}
\end{equation}
The components $g_{ab}$ of the Killing metric can be
defined in terms of the structure constants by the equation
$$
g_{ab} = - {1\over 2n}F^c{}_{ad}F^d{}_{bc}.                  
$$
One lowers and raises indices with $g_{ab}$ and its inverse $g^{ab}$. 

We suppose that $\c{A}$ is a formal algebra with $n$ generators $x^i$
which satisfy commutation relations of the form
\begin{equation}
[x^j, x^k] = i\kbar J^{jk}, \qquad J^{jk} \in \c{A}, 
\qquad (J^{jk})^* = J^{jk}.                                    \label{c-r}
\end{equation}
If the right-hand is considered as given then it must satisfy the
constraints
$$
[x^i, J^{jk}] + [x^j, J^{ki}] +[x^k, J^{ji}] = 0
$$
which follow from the Jacobi identities.  If $J^{ij}$ is
non-degenerate then the center of $\c{A}$ is trivial. The inverse
$J^{-1}_{ij}$ exists in the sense that
$$
J^{-1}_{ij} J^{jk} = \delta_i^k, \qquad J^{-1}_{ij} \in \c{A}.
$$
The algebra has as well $n$ generators $\lambda_a$ which satisfy
the quadratic relations~(\ref{consis}). The commutation relations
between the two sets determines the differential calculus through the
relations~(\ref{frame}).  Consider first the case with $J^{ij}$
central elements of the algebra and with $\lambda_a$ defined by
(\ref{l-x}). This means that $P^{ab}{}_{cd}$ is given by~(\ref{P0})
and that $F^a{}_{bc} = 0$. The associated geometry is flat.  Consider
also the smooth manifold $V = \b{R}^n$ and the algebra $\t{\c{A}}$
generated by the coordinates $\t{x}^i$ and the conjugate momenta
$p_j$. We shall use the convention of distinguishing between the
operator $p_j$ and the result $i\t{\p}_j f$ of the action of $p_j$ on
$f$. There is a simple representation of $\c{A}$ as a subalgebra of
the algebra of (pseudo-)differential operators $\t{\c{A}}$, given by
the identification
\begin{equation}
x^i = \t{x}^i + \frac 12 \kbar J^{ij} p_j.                 \label{deriv}
\end{equation}
From this it follows immediately that
$$
f(x^i) = f(\t{x}^i) + \frac 12 \kbar J^{jk} p_k \partial_j f +
o(\kbar^2) = f(\t{x}^i) + \frac 12 \kbar J^{jk} \partial_j f p_k +
o(\kbar^2)
$$
and from this `Taylor' expansion in phase space we can deduce the
commutation relations
$$
[f \k x^j] = i \kbar J^{ij} \partial_i f + o(\kbar^2)
$$
and hence
$$
[f \k g] = i \kbar J^{ij} \partial_i f \partial_j g + o(\kbar^2).
$$
This can be considered as part of an expression which defines a
noncommutative \nobreak{`$*$-product'} on an algebra of
functions~\cite{Wey50,Moy49} using a formal expression which is an
exponential in the partial derivatives.  If the $J^{ij}$ are not
central then by introducing the vector fields $J^i = J^{ij}
\partial_j$ we can write the commutation relations as
\begin{equation}
[x^i \k  x^j] = \frac 12 i \kbar J^{[ij]} + 
\frac 14 \kbar^2 [J^i \k J^j].                                 \label{rep}
\end{equation}

In this case it is convenient to write~(\ref{deriv}) differently.  We
introduce $n$ vector fields $p_a$ on $\t{\c{A}}$ such that $p_a$ is
the operator which yields $p_a \t{f} = i e_a \t{f}$ when acting on
$\t{f}$ and the $e_a \t{f} = e^i_a(\t{x}^k) \t{\partial}_i \t{f}$ are
the commutative limits of the elements $e_a f \in \c{A}$. We define
also
$$
J^{ij} = J^{ib} e_b x^j = J^{ab} e_a x^i e_b x^j, \qquad
\h{J}^{ij} = \h{J}^{ib} \h{e}_b x^j = 
\h{J}^{ab} \h{e}_a x^i \h{e}_b x^j
$$
and we suppose that $J^{ab}$ is an hermitian central matrix which
satisfies~(\ref{K-J}). Since
$$
e_a J^{ia} = - e_a e_b x^i J^{ab} = - J^{ab} F^c{}_{ab} e_c x^i = 0
$$
the operators $x^i$ are hermitian provided $F^c{}_{ab} = 0$.  This
result relies on the particular form of the product we have chosen
within the algebra of forms.  

If we have two $*$-products as in Section~3 and derivations $e_a$
and $\h{e}_a$ then we can write equivalently Equation~(\ref{deriv})
in the form
\begin{equation}
x^i = \t{x}^i +\frac 12 \kbar J^{ia} p_a, \qquad
\h{x}^i = \t{x}^i +\frac 12 \kbar \h{J}^{ia} \h{p}_a.         \label{eq:1}
\end{equation}
To lowest order this and the perturbed equivalent simplify to respectively
\begin{equation}
[x^i \k x^j] = \frac 12 i \kbar J^{[ij]}(x^i), \qquad
[x^i \hk x^j] = \frac 12 i \kbar \h{J}^{[ij]}(x^i).           \label{s-rep}
\end{equation}
If we define $\theta^{ab}$ by the identities
$$
\h{J}^{ij} = J^{ij} + \theta^{ij}, \qquad 
\theta^{ij} = e_a x^i e_b x^j  \theta^{ab}
$$
then we can write the difference between the commutators as
\begin{equation}
[f \hk g] = [f \k g] + i \kbar  \theta^{ab} e_a f \wm e_b g +  
o(\kbar^2).                                                    \label{W-M}
\end{equation}

In general one would expect that the $\lambda_a$ generate also the
algebra and that each $x^i$ can be expressed as a formal power series
in the $\lambda_a$. The algebra depends then on the coefficients in
the Equation~(\ref{consis}) for $\lambda_a$. In fact the whole
differential calculus depends on these coefficients:
\begin{equation}
\c{A} = \c{A}(P,F,K), \qquad
\Omega^*(\c{A}) = \Omega^*(\c{A})(P,F,K).                    \label{P-F-K}
\end{equation}
We do not imply here that $(P^{cd}{}_{ab},F^c{}_{ab},K_{ab})$ 
are the only parameters. An explicit representation would introduce more.
In the simplest case with $J^{ij}$ a central non-degenerate matrix we
can choose $P^{ab}{}_{cd}$ of the form~(\ref{P0}) and set 
$F^c{}_{ab} = 0$. We find that $x^i$ is linear in $\lambda_a$ and the
relation can be inverted:
\begin{equation}
\lambda_a = \frac 1{i\kbar} J^{-1}_{ai} x^i, \qquad
\h{\lambda}_a = \frac 1{i\kbar} \h{J}^{-1}_{ai} x^i.           \label{l-x}
\end{equation}
We find that $K_{ab}$ is given by the expression
\begin{equation}
K_{ab} = - \frac 1{i\kbar} J^{-1}_{ab},\qquad
i \kbar K_{ac} J^{cb} = - \delta_a^b.                          \label{K-J}
\end{equation}
In this case we can write also
$$
\c{A} = \c{A}(K).
$$
The $\lambda_a$ are represented by
$$
\lambda_a = \frac1{2i} p_a - K_{aj} \t{x}^j.
$$
To a certain extent in this case one might expect that formally at
least the algebra depends only on $K_{ab}$. It is equivalent to a quantized
phase space.  In general we suppose that the commutator is defined in
terms of the $C$-commuta\-tor defined above. That is we write
$$
[x^i,x^j] = [x^i(\lambda_a), x^j(\lambda_a)]
$$
and use (\ref{consis}) to calculate $J^{ij}$ in terms of
$(P^{cd}{}_{ab},F^c{}_{ab},K_{ab})$. In certain cases it might be more
convenient to use a representation of the $\lambda_a$ and from them
construct a representation of the $x^i$ considered as a secondary set
of generators. For example if we set
$$
x^i = i \kbar J_0^{ia} \lambda_a, \qquad J_0^{ib} = \delta^i_a J^{ab},
\quad K_{0,ib} = \delta^a_i K_{ab}
$$
then we find that
$$
[x^i\k x^j] = i \kbar (J_0^{ij} + F_0^{ij}{}_k x^k), \qquad
F_0^{ij}{}_k = F^c{}_{ab} J_0^{ia} J_0^{jb} K_{0,kc}. 
$$
We have here constructed a nonconstant 
$J^{ij} = J_0^{ij} +F_0^{ij}{}_k x^k$ directly from the $\lambda_a$,
which can be considered as comprising the first two terms an an
infinite multipole expansion. More eleborate forms can be obtained by
chossing
$$
e_a \t{x}^i = \delta^i_a + \Lambda^i_a(\t{x}^k).
$$
One obtains then
\begin{equation}
x^i = \t{x}^i + \frac 12 \kbar (J_0^{ib} + 
\Lambda^i_a (\t{x}^k) J^{ab}) p_b.                             \label{eq:3}
\end{equation}
We can choose $x^i$ to be the operator obtained by setting 
$\Lambda^i_a (\t{x}^k) = 0$ and denote $\h{x}^i$ the operator with
generic $|\Lambda^i_a (\t{x}^k)| \ll \delta^i_a$.
Equation~(\ref{eq:3}) can be written as~(\ref{theta}) if we write
$\Lambda^i_a = \delta^j_a \Lambda^i_j$ and set
$$
\xi^i (x^k) = \Lambda^i_j(x^k) (x^j - \t{x}^j).               
$$
Here the variables $\t{x}^a$ are to be considered as parameters. We
deduce, to lowest order, the `Taylor' expansion
$$
f(\h{x}^i) = f(x^i) + 
\frac 12 \kbar (\h{J}^{ab} e_a f \h{p}_b  - J^{ab} e_a f p_b).
$$
If as in Section~3 we write $\h{\lambda}_b = \lambda_b + a_b$ then
from~(\ref{g-c-r}) we find that $a_b$ must satisfy the equation
$$
e_{[a} a_{b]} = \h{K}_{ab} + \lambda_c \h{F}^c{}_{ab} - K_{ab}.
$$
This can also be written as an equality of 2-forms:
$da = d\theta + \theta^2 - \h{\theta}^2$.

The forms $K_{ab}$ and $\h{K}_{ab}$ obviously break Lorentz
invariance, as do the vectors $F_a = \epsilon_{abcd} F^{bcd}$ and
$\h{F}_a = \epsilon_{abcd} \h{F^}{bcd}$. We shall consider these
effects to be of the same order of magnitude as the gravitational
effects. In particular, from this point of view Minkowski space-time
is a degenerate limit. We would prefer to identify the absence of
gravitational field as the commutative limit but it is more convenient
to consider this state as a `regular' cellular structure. The price to
be paid for this assumption is a ground state which is not Lorenz
invariant. This is unfortunate since Lorenz invariance was the
original motivation of noncommutative structure~\cite{Sny47a}.

\setlength{\parskip}{5pt}
\bibliographystyle{utphys}
\bibliography{abbrev,refmad,refgen,proceedings}

\providecommand{\href}[2]{#2}\begingroup\raggedright\begin{thebibliography}{10}

\bibitem{DubKerMad89a}
M.~Dubois-Violette, R.~Kerner, and J.~Madore, ``Gauge bosons in a
  noncommutative geometry,'' {\em Phys.\ Lett.} {\bf B217} (1989)
485--488.

\bibitem{ConLot90}
A.~Connes and J.~Lott, ``Particle models and noncommutative geometry,'' {\em
  Nucl.\ Phys.\ (Proc.\ Suppl.)} {\bf 18B} 29.
\newblock
Annecy-le-Vieux, March 1990.

\bibitem{SchWes92}
J.~Schwenk and J.~Wess, ``A {$q$}-deformed quantum mechanical toy model,'' {\em
  Phys.\ Lett.} {\bf B291} (1992) 273.

\bibitem{Wes99}
J.~Wess, ``$q$-deformed {H}eisenberg algebras,'' in {\em Geometry and
  {Q}cpsuantumphysics}, H.~Gausterer, H.~Grosse, and L.~Pittner, eds., no.~543
  in Lect.\ Notes\ in\ Phys., pp.~311--382.
\newblock Springer-Verlag, 2000.
\newblock {S}chladming, January 1999.

\bibitem{Mad89c}
J.~Madore, ``{K}aluza-{K}lein aspects of noncommutative geometry,'' in {\em
  Differential Geometric Methods in Theoretical Physics}, A.~I. Solomon, ed.,
  pp.~243--252.
\newblock World Scientific Publishing, 1989.
\newblock Chester, August 1988.

\bibitem{Sch99}
V.~Schomerus, ``{$D$}-branes and deformation quantization,'' {\em J.~High
  Energy Phys.} {\bf 06} (1999) 030.

\bibitem{Sny47a}
H.~Snyder, ``Quantized space-time,'' {\em Phys.\ Rev.} {\bf 71} (1947) 38.

\bibitem{DopFreRob95}
S.~Doplicher, K.~Fredenhagen, and J.~Roberts, ``The quantum structure of
  spacetime at the {P}lanck scale and quantum fields,'' {\em Commun.\ Math.\
  Phys.} {\bf 172} (1995) 187.

\bibitem{JurSch00}
B.~Jurco and P.~Schupp, ``Noncommutative {Y}ang-{M}ills from equivalence of
  star products,'' {\em Euro.\ Phys.\ Jour.~C} {\bf 14} (2000) 367,
\href{http://xxx.lanl.gov/abs/hep-th/0001032}{{\tt hep-th/0001032}}.

\bibitem{Ger64}
M.~Gerstenhaber, ``On the deformation of rings and algebras,'' {\em Ann.\ of\
  Math.} {\bf 79} (1964), no.~1, 59--103.

\bibitem{Wey50}
H.~Weyl, {\em The Theory of Groups and Quantum Mechanics}.
\newblock Dover, New York, 1950.

\bibitem{Moy49}
J.~E. Moyel, ``Quantum mechanics as a statistical theory,'' {\em Proc. Camb.
  Phil. Soc.} {\bf 45} (1949) 99.

\bibitem{SeiWit99}
N.~Seiberg and E.~Witten, ``String theory and noncommutative geometry,'' {\em
  J.~High Energy Phys.} {\bf 09} (1999) 032.

\bibitem{MadMouSit97}
J.~Madore, J.~Mourad, and A.~Sitarz, ``Deformations of differential calculi,''
  {\em Mod.\ Phys.\ Lett.~A} {\bf 12} (1997) 975--986,
\href{http://xxx.lanl.gov/abs/hep-th/9601120}{{\tt hep-th/9601120}}.

\bibitem{DimMad96}
A.~Dimakis and J.~Madore, ``Differential calculi and linear connections,'' {\em
  J.~Math.\ Phys.} {\bf 37} (1996), no.~9, 4647--4661,
  \href{http://xxx.lanl.gov/abs/q-alg/9601023}{{\tt q-alg/9601023}}.

\bibitem{MadSchSchWes00a}
J.~Madore, S.~Schraml, P.~Schupp, and J.~Wess, ``Gauge theory on noncommutative
  spaces,'' {\em Euro.\ Phys.\ Jour.~C} {\bf 16} (2000) 161--167,
  \href{http://dx.doi.org/10.1007/s100520000394}{{\tt
  DOI:10.1007/s100520000394}},
\href{http://xxx.lanl.gov/abs/hep-th/0001203}{{\tt hep-th/0001203}}.

\end{thebibliography}\endgroup

\end{document}